\pdfoutput=1
\documentclass[manuscript,prologue,table,xcdraw,nonacm]{acmart} 
\makeatletter
\let\@authorsaddresses\@empty
\makeatother
\setcopyright{none}
\renewcommand\footnotetextcopyrightpermission[1]{} 
\AtBeginDocument{%
  \providecommand\BibTeX{{%
    \normalfont B\kern-0.5em{\scshape i\kern-0.25em b}\kern-0.8em\TeX}}}

\usepackage[utf8]{inputenc}

\usepackage{lipsum}
\usepackage{pgfplots}
\usepackage{amsthm,amsmath}
\usepackage{graphicx}
\usepackage{textcomp}
\usepackage{url}

\usepackage{flushend}

\usepackage[linesnumbered,algoruled,boxed,lined]{algorithm2e}

\newcommand{\ie}{\emph{i.e.}, }
\newcommand{\eg}{\emph{e.g.}, }

\usepackage[makeroom]{cancel}

\usepackage{subfigure}
\usepackage{lscape}
\theoremstyle{plain}

\usepackage{ color, colortbl}
\definecolor{Gray}{gray}{0.9}

\usepackage{tikz}
\usetikzlibrary{arrows, arrows.meta, positioning,shapes,backgrounds,calc,fit}
\usepackage{varwidth}

\definecolor{contC}{RGB}{113,141,165}
\definecolor{techC}{RGB}{129,128,178}
\definecolor{socC}{RGB}{116,174,150}
\definecolor{compC}{RGB}{212,168,106}
\definecolor{compC2}{RGB}{128,83,21}

\tikzstyle{block} = [draw, fill=blue!5, rectangle,     minimum height=3em, minimum width=6em]
\tikzstyle{group} = [draw, rectangle,  minimum height=3em, minimum width=6em]
\tikzstyle{shade} = [ rectangle,  minimum height=3em, minimum
width=6em,inner ysep=2em,inner xsep=0.5em, rounded corners]
\tikzstyle{coord}= [coordinate]

\newcommand{\SY}{\texttt{Sherlock}\xspace}
\newcommand{\B}{\texttt{Bobby}\xspace}

\begin{document}

\title{                                                                  
  Algorithmic audits of algorithms, and the law                                                              
}

\author{Erwan {Le Merrer}}
\affiliation{%
  \institution{Univ Rennes, Inria, CNRS, Irisa}
  \country{France}}
\author{Ronan Pons}
\affiliation{%
  \institution{UT1 Capitole, Université d'Ottawa \& ANITI}
  \country{France \& Canada}}
\author{Gilles Tr\'edan}
\affiliation{%
  \institution{LAAS/CNRS}
  \country{France}}

\date{}

\begin{abstract}

Algorithmic decision making is now widespread, ranging from health care allocation to more common actions such as recommendation or information ranking. The aim to audit these algorithms has grown alongside. In this paper, we focus on external audits that are conducted by interacting with the user side of the target algorithm, hence considered as a black box.
Yet, the legal framework in which these audits take place is mostly ambiguous to researchers developing them: on the one hand, the legal value of the audit outcome is uncertain; on the other hand the auditors' rights and obligations are unclear.

The contribution of this paper is to articulate two canonical audit forms to law, to shed light on these aspects:

\begin{itemize}
    \item the first audit form (we coin the \B audit form) checks a predicate against the algorithm, while the second (\SY) is more loose and opens up to multiple investigations. We find that:
    \B audits are more amenable to prosecution, yet are delicate as operating on real user data. This can lead to reject by a court (notion of admissibility).
    \SY audits craft data for their operation, most notably to build surrogates of the audited algorithm. It is mostly used for acts for \textit{whistleblowing}, as even if accepted as a proof, the evidential value will be low in practice.
    
\item these two forms require the prior respect of a proper right to audit, granted by law or by the platform being audited; otherwise the auditor will be also prone to prosecutions regardless of the audit outcome.
\end{itemize}

This article thus highlights the relation of current audits with law, in order to structure the growing field of algorithm auditing. 


\end{abstract}

\maketitle


Society rules through law, so that law is supposed to hold service providers and their algorithms accountable.
In particular, \textit{decision-making algorithms} are now widespread \cite{10.1145/2844110}. They directly face users, and govern large portions of our lives (from apparently subtle decisions such as recommendations, to more life changing ones such criminal justice or health care allocation \cite{ledford2019millions}). The legal perspective on algorithms, especially of online platforms, is evolving to strong legal frameworks. For example, at an european level, in order to protect the fundamental rights of european residents\footnote{Proposal for a REGULATION OF THE EUROPEAN PARLIAMENT AND OF THE COUNCIL on a Single Market For Digital Services (Digital Services Act) and amending Directive 2000/31/EC COM/2020/825 final.}, or to frame artificial intelligence systems\footnote{Proposal for a REGULATION OF THE EUROPEAN PARLIAMENT AND OF THE COUNCIL LAYING DOWN HARMONISED RULES ON ARTIFICIAL INTELLIGENCE (ARTIFICIAL INTELLIGENCE ACT) AND AMENDING CERTAIN UNION LEGISLATIVE ACTS COM/2021/206 final.} explicitly shows the willingness to better regulate algorithms. 

\paragraph{The IT perspective and the nascent field of audits}
As computer scientists and engineers, we are used to design and develop algorithms that process information and that can have an important impact on society \cite{cacm,10.1145/3442188.3445935}. 
For fine tuning these, developers operate a controlled feedback loop on data fed as inputs to the algorithm, and the corresponding algorithm results (output accuracy for instance).

Considering an exterior viewpoint (the viewpoint of users or regulators), that observes or \textit{audits} the behavior of remote algorithms is less frequent. A so called \textit{black box} approach on algorithms can be dated back to Moore's tests black box automata in 1956 \cite{moore2016gedanken}. Relatively recent and sporadic works instead placed this viewpoint at the service of algorithmic auditing, in order to allow users to gain some understanding on the algorithmic decisions they are facing \cite{xray,aivodji2019fairwashing,10.1145/2844110,buolamwini2018gender, lime, uber, radicalization, curation, remix, fairlens, huszar2021algorithmic, kaiser2019implications}. 
In particular, these nascent forms of algorithmic audits can also constitute a prerequisite to enable platform regulation \cite{10.1145/3469104}: if a state wants to enforce some behavior, means for verification are mandatory (as captured by the Russian proverb \emph{trust, but verify}.

\paragraph{And the law?}
There is a blind spot regarding the current development of algorithmic audit techniques: what is their legal groundings? Little (if any) mentions are made in research works of the legal consequences of the conducted audit. In a nutshell, two fundamental questions are unaddressed: \emph{i)} what are the legal risks taken by an auditor and \emph{ii)} can the outcome of an audit be used against its operating platform in court?
This lack is a structural problem, 
as the destination of the discovered issues must be central.
 Possible recipients of the results of such audits can then be the general public (\eg through the act of \textit{whistleblowing}), or justice. 

In both cases, legal aspects are at stake: the auditor is likely to have violated the terms of service of the audited algorithm on its website. What are then the consequences she faces? Or the consequences for the findings in a trial? If the auditor is in its own right to perfom certain actions, what are the audit steps that will prevent acceptance of a scientific proof in the eyes of justice? %

\paragraph*{Contributions} In an attempt to shed light on the relation between audits and the law, we first propose to bind two prototypes of audit algorithms (that encapsulates state of the art audit algorithms) to specific law perspective. Since the law is by definition a sovereign prerogative to each state in the world, we will take the french law system as an instance and example in our presentation. We will as well include european perspectives from global law frameworks currently in progress.
This yields two salient points: 1) the simplest form of audit (we coin the \B audit) is easily usable in court, yet it is more delicate as it leverages real data (which can be a cause of rejection if all care have not being taken regarding laws such as the \textit{GDPR}: the General data protection regulation\footnote{Règlement ({UE}) 2016/679 du Parlement européen et du Conseil du 27 avril 2016 relatif à la protection des personnes physiques à l'égard du traitement des données à caractère personnel et à la libre circulation de ces données, et abrogeant la directive 95/46/{CE} (règlement général sur la protection des données) (Texte présentant de l'intérêt pour l'{EEE}).}. 2) The most complex audit form (we coin the \SY audit) is less problematic as it crafts data as inputs; yet, it is far more difficult to bring to justice due to possibly lower probative value, leaving it in priority for whistleblowing. Finally, we review the conditions for an audit to fit in the whistleblower category.

\section{Two canonical algorithms capturing algorithmic audit schemes} 
\label{sec:algorithms}

There is a growing diversity of audits in the recent literature \cite{xray,aivodji2019fairwashing,10.1145/2844110,buolamwini2018gender, lime, uber, radicalization, curation, remix, fairlens, huszar2021algorithmic, kaiser2019implications}. Each one spans a specific behaviour of a specific platform with its own methodology. In an attempt to structure this nascent field we introduce a set of fundamental distinctions that allows to separate those audits into two broad categories that distinguish audits both on their technical approach, and on their relevance for a potential trial.  We introduce each category through an archetypal algorithm, that is an abstract high-level representation of the audits it describes. We then showcase how some concrete audits of the literature fit each archetypal algorithm.

To introduce our categories, let us take the parallel with police work, tasked to check the application of law. On one hand, \B-family audits are tasked to tour the audited algorithm evaluating a well defined characteristic of the platform, similarly to a policeman tasked to tour a district to fine car parking infringements. Key to this approach is the existence of a logical predicate that very precisely defines the desirable (resp. undesirable) behaviour of the audited algorithm, similarly to the set of driving regulations that precisely define what is a correctly parked car. 
On the other hand, \SY-family audits target a deeper and loosely defined characterization of some aspect of the audited algorithm, similarly to an inspector trying to elucidate some crime. Such approach typically requires some interpolation in order to provide a general analysis based on some observed examples of the algorithm behaviour.

The basic material audits are built on are algorithm outputs, corresponding to inputs the auditor has submitted.

\subsection{Context - Terminology}

The use case we consider is the following: an individual (or a group of) hereafter named \emph{the auditor} seeks to study the behaviour of some algorithm executed remotely by some platform hereafter named \emph{target platform} or the \emph{target algorithm}. We focus on the case where auditors are completely external to the target platform, and hence can only interact with the public side of the algorithm as would regular users do. 

This context is tailored to represent the typical context of a black box audit, where auditors are simple users interested in understanding or evaluating the behaviour of the platform they use. This context can also capture situations in which the competent authorities have no specific access to the algorithm
and wants to verify the compliance of this behaviour with some regulation. 

Concretely, we wish to capture a spectrum of use cases ranging from informal citizen-driven audits (see \eg COMPAS) to academic research work on platforms. All those situations cover the same high level steps: an auditor writes some code to \emph{i)} request the target platform (either through some API, either through its web interface directly), \emph{ii)} parse and collect the target algorithm answers and \emph{iii)} publicises some analyses based on the collected data.

Formally, let $A$ be the target algorithm. Let $X$ (resp. $Y$) be the input (resp. output) space of $A$. Like regular users, auditors can only submit some request $x\in X$ to $A$, and then record the corresponding result $A(x) \in Y$.

\subsection{The \B audit form} 

This is the simplest category of audit algorithms.  
In this audit form, an infraction is constituted by an input set (that is a data existing in a dataset), to which corresponds a (problematic) collected output.

\begin{algorithm}
\small
\KwIn{$A$ an algorithm to audit. $A:X\mapsto Y$\\
$L$ a propositional formula over an input dataset  $X_L=\{x_1\ldots x_l\}\subset X$ and corresponding outputs $Y_l=\{A(x_1),\ldots,A(x_l)\}\subset Y$
}

\KwOut{\emph{True} if the behaviour is illegal, \emph{False} otherwise}

infraction = False\\

\For{$X$ not exhausted}{
    Pick $x_1,\ldots,x_l$\\
    Collect $Y_1 = A(x_1),\ldots,Y_l = A(x_l)$\\
    \uIf{ \textbf{not} $L(X_1,Y_1),\ldots$ or $L(X_l,Y_l)$}{
        infraction = True  \tcp*{$A$ does not verify $L$} 
        \textbf{break}
    }
}
\Return infraction \tcp*{Boolean on violation of $L$} 

\caption{\small The \B audit}
\label{alg:bobby}
\end{algorithm}

\paragraph{Propositional formula $L$}
In the pseudo-code presented on Algorithm \ref{alg:bobby}, the central component is the definition of propositional formula $L$ to be checked against the audited algorithm. In its definition, $L$ encodes the desirable property one wants to observe. More precisely $L$ is a propositional formula defined over a set of input/output couples of the target algorithm that constitute the variables of the proposition. In $L$, those variables are linked by logical operators such that $L$ is well formed and has a \emph{truth value}: $L$ is either true or false.  

As an illustrating example, imagine $A$ is the algorithm that is in use in an online flight search platform that allows users to seek and book flights. 
For each request (departure and destination locations belonging to the IATA list, and dates), it provides the user with an ordered list of flights $f_1,f_2,\ldots$. Assume the platform operating $A$ declares that it ranks the resulting flights according to their cost. Such assertion can be easily converted to a propositional formula that can be evaluated over any couple of flights $f_i,f_j$: $L_{cost}(f_i,f_j):= i\leq j \Rightarrow cost(f_i) \leq cost(f_j)$. Such declaration can be audited with a \B audit that regularly requests $A$ to verify if $L$ holds. In our example, an input for which $L$ is violated is a couple of two returned flights $f_a,f_b$ such that $f_a$ is more expensive and yet presented before $f_b$ (formally: $a<b \wedge cost(f_a)>cost(f_b)$). If such violating input is found, the algorithm stops and reports the observed behaviour.

\subsubsection{ \B approaches in the literature}

We now illustrate some concrete \B examples available in literature.

\paragraph{Cookies/ Transparency Consent Form auditing}
The GDPR and ePrivacy Directive recently set that European users must explicitly consent to non-necessary data collection, in general stored as a consent cookie on the users' computers. This rule can easily be automatically audited, \cite{matte2020cookie} implemented a crawler that \emph{i)} visits a target website without any interaction and \emph{ii)} detects the writing of a cookie registering consent by the target website. In such typical \B audit, input space could be $X=$ all the target's webpages, and the predicate is in this case simple: $L= $not positive consent cookie.

\paragraph{The detection of "fairwashed" explanations.}
Online service are now increasingly proposing to explain the main factors driving some of their automated decisions.
The rationale is for them to gain trust by the general public. Nevertheless, there is a possibility that the provided explanation are faked (fairwashed \cite{aivodji2019fairwashing}) to justify a discriminative decision. In \cite{LMT20-nat}, so called incoherent pairs are looked for; these are two conflicting explanations that yet give the same decision, and are the sign of a fairwashed explanation by the audited algorithm.
This can be written as $L=\{ \nexists In=((a,white),(a,black)) \in X^2~s.t.~ f(a,white) \neq f(a,black)\}$. 
This mimics the work of associations that are performing tests at the entrance of clubs for instance.

\paragraph{Copyright or backdoor auditing}
Some forms of audits are to potentially identify a remote algorithm that is infringing some copyrights (by being executed without permission). The audit result is Boolean answer on whether or not the remote algorithm is indeed the one that is suspected.
This relates to the field of \textit{watermarking}, where an algorithm is queried, and returns specific outputs if it is indeed the one suspected of infringement \cite{usenix-wat}. 
Here, the inputs used as queries are specifically designed to operate as identification keys for that purpose.
The predicate resemble $L=\{\forall In \in K_g, f(In)==g(In)\}$, with $K_g$ being the watermark \textit{key}.

\paragraph{Skin color or gender bias audits.}
Multiple studies fit in this class: to take a precise example, Buolamwini et al. \cite{buolamwini2018gender} benchmark three commercial gender classifier systems with an intersectional dataset (skin color/gender). Gender classification accuracies are compared: the paper notes that \eg classification is $8.1\%$ to $20.6\%$ worse on female than male subjects and $11.8\%$ to $19.2\%$ worse on darker than lighter subjects.

Interestingly, this paper first constructs a dataset of faces that has balanced gender and skin types. Assuming this dataset is standardly recognised as a good benchmark for face classification, one could imagine a \B approach that targets any face classification algorithm using as input $D:$ the standardized dataset. 
To implement the predicate function from Algorithm~\ref{alg:bobby}, classification results obtained on $D$ could be compared for instance against a $60\%$ disparity ratio \cite{feldman2015certifying}. For any partition of $D$ into a gender/skin type subset $D_s$ and its complementary $D_{\bar{s}}$, one has to compute the target algorithm's accuracy: $a_s=1/|D_s|.\sum_{i,label(i)\in D_s} A(i)==label(i)$. For each partition $s$ covered by the dataset, one then evaluates the predicate $L_s=\frac{a_s}{a_{\bar{s}}} > 0.6$.

\paragraph{Diversity in search engine results}

Urman et al. \cite{urman2021auditing} track several search engines, to audit source diversity and search concentration. This is achieved by submitting a static list of 62 keywords. Like for bias, the final predicate can take the form of a simple rule such as one where the diversity at search engine B must be at least $0.6$ the one at B for instance.

\subsubsection{Limits of \B} 

The \B forms of audits are bound to verify a predicate $L$ over an input space $X$. Given an input budget $N$ (\ie the amount of different input queries sent to $A$), three outcomes are possible. Among them, two are to be considered  as potential limitations.

Either some input $c\in X$ violating $L$ is found (ie $L(c)$ is false). In this case, an infraction has been found, and simply exhibiting $c$ and its corresponding answers $A(c)$ is sufficient to establish the infraction to $L$ committed by $A$.

Either no input violating $L$ was found. We need to distinguish 2 sub-cases: 
        \begin{itemize}
            \item $ N > \vert X \vert$: the whole input space of $A$ is exhausted, and no violation has been found. It is then legitimate to conclude that $L$ is respected by $A$. Unfortunately, current algorithms have input spaces that are either unbound (\eg with inputs being floats) or have a size orders of magnitude larger than typical auditing budgets $N$ (\eg few hundred of queries for an input size of $3 \times 224 \times 224$ corresponding to images \cite{maho2021surfree}).
            \item $N < \vert X \vert $: no violation of $L$ has been found within the budget $N$. In this quite common case, the auditor is left with a statistical guarantee but no definite answer. While the precise nature of the statistical guarantee depends on the specifics of the study (\eg how the input space is sampled using the available input dataset), such an assertion typically translates that the empirical probability $\hat{P}_{\overline{L(c)}}$ of finding an input $c$ violating $L$ is less than $1/N$.
        \end{itemize}

A second limitation of \B resides in the production of inputs to query the target algorithm. We here stress that inputs from $X$ (l. 2) can belong to a dataset (\eg image dataset in \cite{buolamwini2018gender}), or be formed by public data. Thus, this line hides a wide variety of situations that are both heterogeneous with respect to the technical difficulty of generating $X$ and heterogeneous with respect to the legal consequences.

Regarding the technical difficulty first: target algorithms working on simple inputs, like a text request on a search engine, might be queried with datasets $X$ that are easily collected. On the other hand, target algorithms working on more complex inputs, like a video, a resume or a medical record, might face the auditor with a greater challenge for constituting $X$.

A similar legal heterogeneity also resides in constituting $X$. Consider a flight search engine whose input is constituted by airport names along with some future travel date: both are public data that the auditor can rightfully use. On the other hand, testing a job recommendation engine that would match candidate resumes with job offers might require the auditor to submit resumes found on the web for which privacy rights (among others) exist and requires specific conditions to be processed (i.e. consent of the data subjects). In this second case, a judge might consider that the auditor had no right to use these data, and hence refuse to consider results obtained with it.

A third limitation of \B is the reliance on a propositional formula $L$: while some desirable behaviours of target algorithms can easily be converted into propositional formulas (\eg if declared age is below 9, do not show ads), some others are intrinsically impossible to convert to such a logical statement (\eg if declared age is below 9 do not propose shocking videos). This limits the applicability of \B audits to some specific and well defined properties of audited algorithms.
For all other cases where the target property is not as defined, a more elaborate approach is required.

\subsection{The \SY audit form} This second form of audit algorithms is more flexible and do not focus on the verification of a single propositional formula. These audits target a different set of infractions that,   instead of relying on the collected outputs alone, rather relate to the general behaviour of the audited algorithm. 
  To pursue the parallel with policemen: building the case for, say, a murder
  requires our policeman to
  come up with a complete narrative (including motives, absence of alibi, etc.) that typically cannot be covered by a single propositional formula. 
  
  A \SY audit also needs to collect interaction sessions that characterize the target behaviour (\ie sequences of input and output pairs), and based on these examples, to \emph{interpolate} on the behaviour of the target algorithm. A typical \SY audit thus contains two phases: a first phase that builds a local model of the target algorithm (hereafter named a \emph{surrogate}), and a second phase that analyses the surrogate to extract its desired properties. 


The algorithm is presented in Algorithm \ref{alg:scotland}.
The input crafting operation (l. 3) is here central: it pertains to a general plan to extract specific information from $A$, in order to create an accurate surrogate for $A$, on the auditor's machine. This local surrogate $S$ is then analysed locally. While \B audits simply evaluate a predicate $L$, the analysis of $S$ is much more open-ended (ranging from identifying shocking corner cases to characterizing the internal logic and comparing against other $A$ equivalents). To capture this diversity in a compact way, we define the set \emph{Acceptable} of situations the auditor would refer to when conducting such an analysis.

\begin{algorithm}
\small
\KwIn{$A$: an algorithm to audit. $A:X\mapsto Y$\\
 $N$: A budget (maximum number) of queries}
I $\gets$ find input to $A$\\

\tcc{Build a surrogate $S$ to $A$}
\While{$n<N$}{
  craft a new $I_n \in$ I\\
  interact with $A$ through $I_n$; collect $A(I_n)$\\
  $S \gets Retrain(S \cup {(I_n, A(I_n))}$ \\
  }
\tcc{$S$ is now a constructed surrogate of $A$. Analyze $S$}
\Return $Analysis$(evidence) is Acceptable \tcc{Return false if some violations are found}

\caption{\small The \SY audit} 
\label{alg:scotland}
\end{algorithm}

As an illustrating example, consider the same online flight search platform, driven by algorithm $A$. Let us in this case assume the target platform does not declare anything on the techniques $A$ relies on to rank flights $F(c)=f_0,\ldots,f_j$ (maybe merely using the term "relevance"). A typical \SY audit task would be to study and understand how $A$ ranks its results. Hence, an approach here could consist in collecting many example rankings $F(c_1), F(c_2), \ldots$, and study without any prior different factors (cost, but also duration, number of layovers, departure time, company) that could explain (correlate) with the ranking of flights. 

A typical use case for such task would be to show that $A$ deliberately favours the flights of some company they are in business with. This example relates to the historical case of SABRE, American Airline's flight reservation system, that used "screen science" to favour its own flights over its competitors by systematically presenting competitors on the second page of the search results \cite{sabre}. 

With \SY, and as opposed to \B, the input data can be fully crafted. This means that there is no prerequisite for a dataset; data can be forged with the objective of triggering some specific behavior for the remote algorithm.
This is precisely what line 5 in Algorithm \ref{alg:scotland} builds on: new input/output pairs are used to retrain a surrogate, that will become closer and closer to the audited algorithm.

\subsection{Reduction to \SY}

We now list a set of notorious research works that fall into the \SY audit form.


\paragraph{Surge price forecasting for Uber}

We refer to the paper entitled "Peeking Beneath the Hood of Uber" \cite{uber}, where authors 
rely on some data capture (measured supply, demand, estimated waiting times and surge prices) to fit three linear regression models. Their aim is to predict the surge multiplier in the next 5-minute interval. 
The inputs are crafted from using several smart phones, for in particular bringing a variety of locations in these inputs. 

This fits directly the core of Algorithm \ref{alg:scotland}, where a surrogate is trained from the queried data, so that after the query budget is over, the surrogate is used to perform a final test.

\paragraph{Tracking action consequences in outputs with XRay}

The audit in \cite{xray} creates fake accounts  to make them interact with the audited platform (Gmail for instance), and detect which data input (\eg email) have likely triggered a particular output (\eg received ad in Gmail).
Distinct ads on each account are tracked. A correlation engine is run, in order to associate inputs and outputs.
To that end, the placement of inputs on given accounts is crucial to be able to properly infer associations.

A Bayesian model is proposed as a surrogate to simulate the audited service given some targeting associations. This is done by computing probability to observe certain outputs depending on targeting associations. 


\paragraph{Explaining ML decisions with LIME}

The goal of LIME \cite{lime} is to explain the decisions of a remote ML model in the vicinity of a given input $x$, by training sparse linear surrogate models as explanations. 
Input samples are drawn uniformly at random around $x$, to obtain a perturbed dataset (along with its labels returned by the remote model). 

\paragraph{COMPAS}

Another notorious example of such an audit form is the analysis of COMPAS\footnote{COMPAS stands for "Correctional Offender Management Profiling for Alternative Sanctions". About the 2016 analysis: \url{  https://www.propublica.org/article/how-we-analyzed-the-compas-recidivism-algorithm}.}, an algorithm used by judges, probation and parole officers to assess a criminal defendant’s likelihood of becoming a recidivist.
This study was used to whistleblow on the bias present in the audited models.

\subsubsection{Limits of \SY} 

While \SY audits can be in principle  target any algorithm, it comes at a price: first a greater cost, both the amount of human intervention required to exploit obtained results and in the amount of requests such results usually require. Second a weakened power as the conclusions of the audit are ultimately drawn from a model whose interpolating power can always be questioned.

\SY audit forms first usually require more human intervention in their design and exploitation. This can first be explained from a purely computational perspective: \B audits are bound to extract a binary information from the target (namely, $L$ is true or false); hence literally extracting the minimal amount of information, while \SY audits are supposed to extract much more information. Consider for instance the airline ranking audit: while the \B information only verifies the statement issued by the platform, \SY are supposed to come up with a narrative identifying the behaviour of the target through generalising a handful of observations. Typical involved steps are: identify potential functions corresponding to the observed behaviour, test and validate potential functions, confirm or infirm each one, confirm conclusions drawn on surrogate with real target, and so on. Such steps are time-consuming, and involve a wide variety of skills such as exploratory analysis, statistics, and literature review.

The second price is the number of requests. Indeed, training a local surrogate model naturally exhibits a trade-off between the training set size (\ie the number of requests issued to the target algorithm) and the accuracy of the resulting surrogate. Hence, to achieve good accuracy, large volumes of input data are often required. As a result, means to automate the generation of inputs to query the target are often necessary. A positive side effect is that generated data will not be protected (unlike personal data). However, such automation might not be easy, or require considerable human intervention.

\begin{figure*}
    \centering
      \pgfdeclarelayer{bg}    
  \pgfsetlayers{bg,main}  

  \begin{tikzpicture}[node distance=2cm]
    \node[block, text width=4cm,label={Audited Platform}] (auditee) {\vbox{
        \begin{itemize}
          \setlength\itemsep{-.3em}
        \item Data nature: privacy, availability
        \item Data requirement: Input Space size
        \item Target Algorithm: offense type
        \end{itemize}
      }};

    \node[block, right of = auditee, node distance=6cm,text
    width=4cm,label={Auditor Entity}] (who) {
      \vbox{
        \begin{itemize}
          \setlength\itemsep{-.3em}
        \item Authorisation to audit ; work-related to the audited algorithm owner ; Concerned by audited algorithm services
        \item Power/Ressources
        \item National legislation
        \end{itemize}        
      }
    };

    \begin{pgfonlayer}{bg}
      \node[shade,fill=contC!60,fit={(auditee) (who)},label={\textsc{Context}}, inner
      ysep=2em] (context) {};      
    \end{pgfonlayer}

    \node[coord, below of= context,node distance=6.5cm] (contextb) {};
    
    \node[draw, circle, fill=compC!60,thin,right of= contextb, node
    distance=3cm] (strategy) {Strategy};
    
    \node[right of= strategy, text width=2cm,coord] (whistle)    {};

    \node[block,text width=5cm, above of= whistle, node distance=2.5cm,label={Judiciary Action}] (legal) { \vbox{
        \begin{itemize}
          \setlength\itemsep{-.3em}
        \item Notifying audited entity
        \item Presence of stake
        \item Loyalty of proof
        \end{itemize}}};

    \node[block,text width=5cm, below of= whistle, node distance=2.5cm, label=below:{Public Disclosure}]
    (public) {\vbox{
        \begin{itemize}
          \setlength\itemsep{-.3em}
        \item Leverage public opinion
        \item Arguments can take any form
        \item No impact guarantee
        \item Legal exposure
        \end{itemize}}};

    \begin{pgfonlayer}{bg}    
      \node[shade, fill=socC!60,fit={(legal) (strategy) (whistle)
        (public) (whistle)},label={\textsc{Societal Dimension}}] (society) {};
    \end{pgfonlayer}    

    \node[draw, circle, fill=compC!60,thin,left of= contextb, node
    distance=3cm] (approach) {Approach};

    \node[left of= approach, coord] (appcoord) {};
    \node[block,text width=5cm, above of= appcoord, node distance=2.5cm,
    label={\B}] (bobby) {\vbox{
        \begin{itemize}
          \setlength\itemsep{-.3em}
        \item Clear predicate
        \item Intelligible proof
        \item Partial coverage
        \end{itemize}}};

    \node[block,text width=5cm, below of= appcoord, node
    distance=2.5cm,label=below:{\SY}]
    (sy) { \vbox{
        \begin{itemize}
          \setlength\itemsep{-.3em}
        \item Surrogate-based approach
        \item Constructed representativity
        \item Importance of assumptions
        \end{itemize}}};

    \begin{pgfonlayer}{bg}    
      \node[shade, fill=techC!60,fit={(bobby) (sy) (approach)},label={\textsc{Technical Dimension}} ] (tech) {};
    \end{pgfonlayer}    


    \begin{pgfonlayer}{bg}
      \draw[dashed, thin] (legal.-25) to[bend left] node[above,rotate=-90,
      text width=2cm,align=center, draw] { \emph{WhistleBlower Procedure}} (public.30);
    \end{pgfonlayer}
    \draw[ultra thick, compC2,->] (auditee)  -| node[above,sloped,pos=.7]
    {\textbf{Determines}} (contextb)  -- (approach);
    \draw[ultra thick, compC2,->] (who) -| (contextb)  -- (strategy);

    \draw[thin,->] (strategy) -| node[pos=.6,rotate=-90,align=center,above]
    {Audit \\Destination} (legal);
    \draw[thin,->] (strategy) -| (public);

    \draw[thin,->] (approach) -| node[pos=.6,above,rotate=90,align=center]
    {Audit\\Implementation} (bobby);
    \draw[thin,->] (approach) -| (sy);

    \draw[ultra thick,teal,->,pos=.35] (bobby) -- node[text
    width=1.5cm] {Intuitive Approach} (legal);

    \draw[ultra thick,teal,->,pos=.25] (sy.east) -- node[sloped,text
    width=1.5cm] {Complex Approach} (legal);

  \end{tikzpicture}

    \caption{Overview of the perspectives presented in this article: a given situation (auditor, target algorithm and offense) defines both possibilities in terms of legal outcomes and in terms of auditing approaches. In favourable situations, law defines a clear predicate that can simply be tested online. When no such predicate exists, auditors have to infer more about the target. Relying on such inferences in a court case is more complex, hence those audits are often used to leverage public opinion through whistleblowing.}
    \label{fig:overview}
\end{figure*}
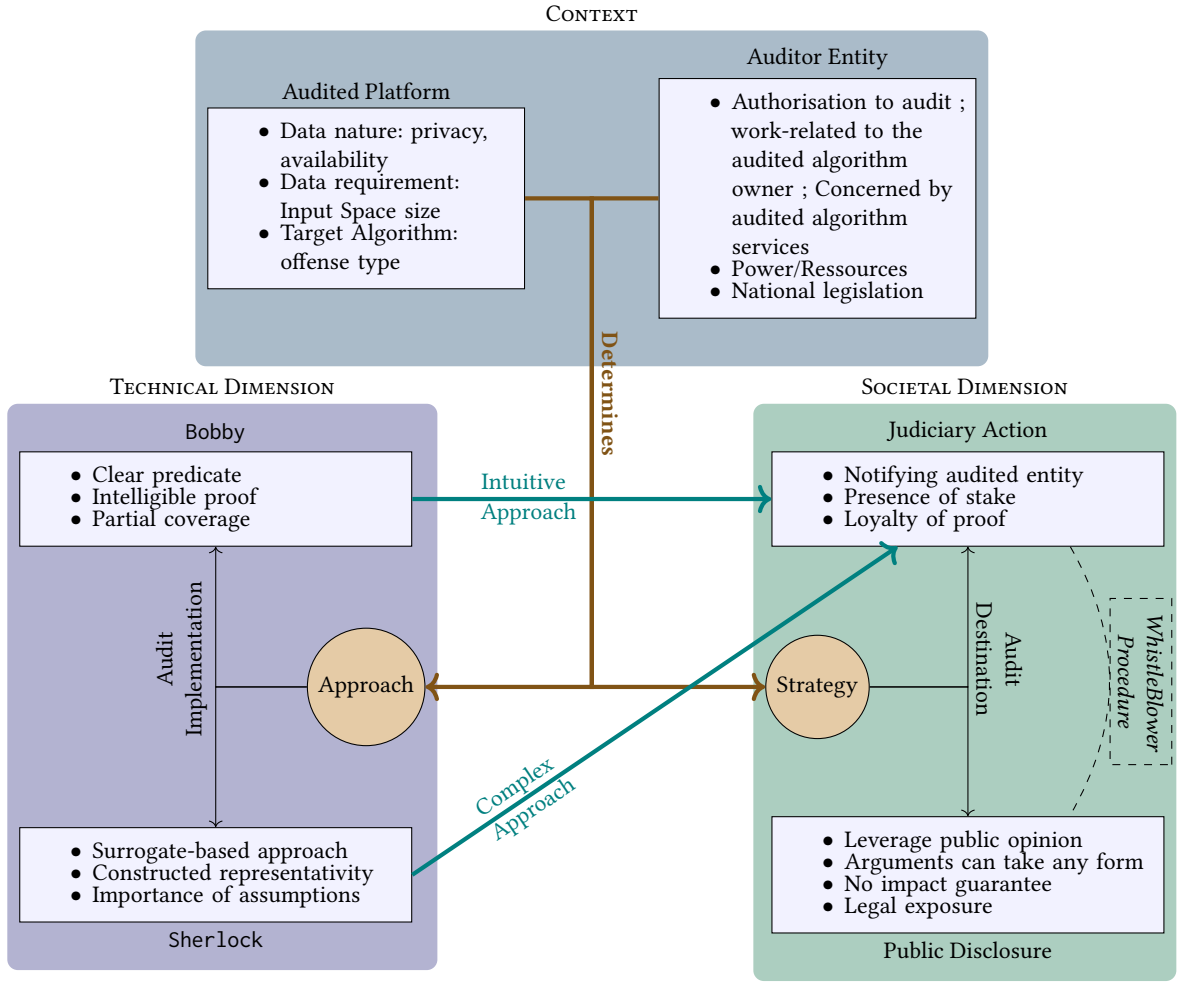

\section{The law perspective on audits}

We now discuss the interplay of audits with the law (as summarized in Figure \ref{fig:overview}).

\paragraph{Disclaimer.}
The following part is dealing mostly with a specific field of the law called \textit{procedural law}. Procedural law refers to the rules of judicial organization, jurisdiction, trial proceedings and enforcement of court decisions, including administrative, civil and criminal proceedings. We are going to focus on the french rules of legal proceeding, which will be different in another country.
Even within a country, these judicial proceeding rules differ between the situation you are facing. Indeed, administrative litigation will not be ruled by the same provisions than a criminal litigation or a civil one (we will address the latter). 
If some of the rules also applied to the administrative case or penal case in France, it will be mentioned explicitly. 
We stress that we do not intend to be exhaustive about the legal proceedings rules, but to help providing insights about the legal context for their specific situations in auditing an automated decision making algorithm.

\subsection{Legal issues regarding the auditor}

The legal situation of auditing a target algorithm depends significantly on the identity of the auditor. Firstly, the most important difference between two auditors is whether they have the authorisation to realise the audit. A regulatory authority, commissioned by the law to evaluate the compliance of online algorithms to specific regulation, profits of specific powers. Powers that enable the authority to control companies' online algorithms without facing most of the legal risks encountered by a regular person. Likewise, a contract between the auditor and the audited algorithm operator allows the auditor to realize some actions which would represent a violation of legal terms for others people. 
Secondly, the relationship between the auditor and the audited algorithm also plays an important role on which ways audit results can be used. The "standing" is a legal term referring to the existence of an interest in the claim for the claimant\footnote{Legal requirement provided in article 31 of the french code of civil procedure.}. In other words, the claimant (the auditor in our context) must gain a benefit or avoid a loss through the court action. Otherwise, her action will be denied by the judge\footnote{Article 32 of the french code of civil procedure.}. However, having the possibility to bring a claim to court does not give the auditor any right to audit the controversial target algorithm without a proper authorisation. 
 
Identifying the nature of the auditors compared to the target algorithm (a customer, worried to loose money, loan applicants, suspicious about the online algorithm in charge of grants or just a individual who found out a illegal situation) is an important prerequisite to audit.

\subsection{Legal issues regarding the inputs used to audit}

From a legal perspective, a clear split can be performed based on how the inputs used by the two audit forms are chosen. While \SY fully crafts its inputs for the purpose of its investigations, \B uses existing data (\ie existing pictures or user profiles) as a basis for its audit.

Regarding that matter, the difference is really significant. Nowadays, existing data of all types can be protected by plenty of legal texts. The GDPR presents the rules of protection on personal data\footnote{GDPR defines \textit{personal data} as "any information relating to an identified or identifiable natural person" which is an extremely wide definition.}. Every action made on personal data is called \textit{processing}"\footnote{Article 4 of the GDPR defines processing as "any operation or set of operation which is performed on personal data", e.g. collection, recording, consultation, alteration, use,etc.} these data. Furthermore, there are special categories of personal data which are subject to additional protections to be processed. 
Data can also be protected by intellectual property law and/or by database protection. In this context, european scientists are benefiting (depending on when their country is adapting the european directive on copyrights\footnote{Article 3 of DIRECTIVE (EU) 2019/790 OF THE EUROPEAN PARLIAMENT AND OF THE COUNCIL of 17 April 2019 on copyright and related rights in the Digital Single Market and amending Directives 96/9/EC and 2001/29/EC.} into the national legal framework) from a data mining exception for research (and non-commercial) purpose. This protection applies when a lawful access to the data was performed; otherwise audit actions could be considered as an infringement to intellectual property. 
Those are just two examples of the many regulation that can be applied on data. From health data to passengers data, processing them without complying to the legal obligations associated can expose to important economic and criminal sanctions\footnote{For instance, processing personal can be sanctioned by administrative fines, upon 20 millions euros or 4\% of the total revenue. French criminal law also provides a sanction of 5 years of imprisonment and a 300.000€ fine for people collecting personal data by fraudulent, unfair or unlawful means.}. 

We now warn about the counter intuitive consequences of using auditing algorithms. \B audits, which looks easier to design than \SY audits, expose the user to more legal risks because of the processing of existing data. The complexity of the audit algorithms used is not correlated with the amount legal issues it faces.

\subsection{Scientific proof vs Legal evidence}

\subsubsection{Admissibility of the proof} 

Before going to court, an auditor must inform the target algorithm platform of the problematic elements they have found. The Digital services act (DSA) proposal wants an obligation to put in place mechanisms in order to allow any individual or entity to notify any potential illegal content hosted. Even now, this mechanism is already existing on a lot a websites. Collaboration with the targeted platform remains the fastest way to remove the controversial content found through an audit. 

When scientists realize an investigation or an experiment, they produce results which will be evaluated by their peers within their research community. The results are evaluated depending on the criteria of the community and this will decide the impact on the field.
Nevertheless, the acceptance by a scientific community does not guarantee for the proof or results to be considered by the judge during a case. 

Law has its own criteria when it comes to the admissibility of legal evidences. It is a specific discipline in law studies called \textit{procedural law}. Depending on the legal system the auditors are operating in, those can be written explicitly or not. 
In France, the admissibility of an evidence depends on the loyalty of the establishment of the evidence. The "loyalty of proof" principle is a recurring principle in the french legal system. However the specification of this principle will change with the field of law the auditor is working in. As an example, the loyalty of proof principle is different in criminal law and in administrative law, and different in civil law. Even within these legal domains the principle could differ from one action to another. Law is all about contexts and exceptions. 

In particular, the legality of the evidence principle is described in the french Civil Code\footnote{Article 9 : "Each party has the burden of proving in accordance with the law the facts necessary for the success of its claim."}. It means that an evidence  obtained through an illegal manner cannot be used in court afterward. For instance, an employer who is using personal data as camera recording without informing the employees is contrary to the GDPR obligations. Therefore, the video surveillance recording will be refused as legal evidence by the judge. The technical proof will not be accepted, before any consideration to its quality as a technical element. Elements found in a computer or any IT system without the proper authorization\footnote{About the sanctions of fraudulent access or maintain in a IT system, see art 323-1 of the french penal code.} to do so will not be accepted by the judge, even with proofs of the defendant guilt. Indeed, a piece of evidence obtained by means of an unfair process is inadmissible. Some exceptions to this principle exist in criminal law or in labour law. 

However the acceptation of the evidence relies on the interpretation of the judge. {\bf The infringement(s) must be \textit{necessary and proportionate}}\footnote{These notions of necessity and proportionality of a legal evidence has been admitted in first place by the European court of justice. For more information, see J. Van Compernolle, « Les exigences du procès équitable et l’administration des preuves dans le procès civil », RTDH 2012. 429.} to the purpose, \ie finding evidence to support a claim. A necessary evidence means the illegal or disloyal evidence brought by the claimant is their only solution to support their claim. In other words, was the infringement or disloyalty necessary to support the claim ? Then the judge will perform the proportionality test. She will balance between the opposed side rights and liberties which have been violated by the audit establishing the evidence and the evidence stakes for the claimant. This analysis of the necessity and the proportionality is solely based on facts. It means no generalization can be made out of the decisions given by the judges already\footnote{G. Lardeux, « Le droit à la < preuve > : tentative de systématisation », RTD civ. 2017. 1.}. In the end, if the judge qualifies the proof as necessary and proportionate, the evidence will be admitted in court. 

This concern echoes with the inputs used by the two forms of audits. When someone is auditing without proper authorization, not only the author get responsible for the violation of rights realized during the process, but also she is making the admissibility of the proof in court more difficult. If the audit purpose is to sue the provider afterward, one has to be extremely cautious about the violation realized during the audit. Otherwise the potential legal consequences of an investigation will be jeopardized.

It is clear that in most situations, getting the right elements is impossible without a violation of rights, either derived from the law itself or a contract\footnote{an infringement of terms and conditions of use of a website. For instance, YouTube' terms of service explicitly specify that accessing their services through automated processes is not allowed (except for specific situations).}. It does not mean that the condition of \textit{necessity} is fulfilled. There can be a significant difference between two violations of rights. The severity of the situation relies on multiple factors as the right or rights violated, the rights owners' quality, the quantity of infringements realized for the audit, etc. Designing an auditing algorithm which makes the less severe violation to the law is important to increase the chances of admissibility by satisfying (at least theoretically) the condition of necessity. The fact that a better suited audit algorithm could have been used to search for the targeted element, will reduce the chances to see the evidence being admitted in court.

The \B and the \SY audit forms serve different purposes. The choice between each the two must be thought with caution. After the choice of a certain type of audit, the choice of actions performed within it is evenly important. 

\subsubsection{Probative value of the audit results}

We have discussed about the admissibility of the scientific proof in the court. But {\bf being admissible on the court is not related to the level of importance given to the scientific proof}. Legal rules distinguish between the admissibility of the evidence and its evidential value. Admissibility means whether or not the elements will be accepted in the list of evidences, while the value of the proof means the value given in the court by the judge of the evidence that have been accepted beforehand. In civil law, the probative value of some evidence is written explicitly in the law. In that case, they are binding on the judge who cannot therefore evaluate their credibility on her own conviction\footnote{G. LARDEUX, "Preuves : modes de preuve" (2019), Répertoire de droit civil.}. Algorithmic results do not belong to this category: their evidential strength will be sovereignly assessed by the judges. 

Irrespective to the form of audits one is intending to use, there is one limit faced by all audit algorithms which are trying to discover illegal situations. Their target is a \textit{technical representation} (\ie a branching on a decision, a weight on a profile variable, \dots), and this is very different from the language used in laws. 
Translating unclear terms of a legal text enforces an interpretation from the auditor. This interpretation is subjective and may not be adequate to the state of law and interpretation of the judge. Audits will succeed more in identifying simple and explicit illegal situations, which are clearly defined in legal texts.
For this kind of situations, the output probative value can always be challenged on the basis that it does not represent what the law definition meant on the first place: this is an inherent limit to all audits to be aware of before taking legal risks, possibly infringements, in the audit of an algorithm.

\B audits provide results which exist in the environment of the audited algorithms. The outputs are retrieved when sending inputs, but not created per se by the audit. \B  allows for automatic research through a digital environment which could have been done manually (even by a lot of persons tasked to do so). It means that the result can be found again by a person if a confirmation of the output existence is needed. There are one positive and one negative legal consequences regarding this statement of "existing results". 
Starting with the negative one, the \B audit are useful in court only when manifest evidence are needed. By "manifest" evidence, law refers to illegal content or situation which are noticeable. There is no need for further justification for the claimant.
For example, there are some abusive terms in consumer contracts which are specifically forbidden by the french consumer law. When this kind of abusive terms are found, the professional must withdraw them from the contract without any challenge available. 
The utilisation of \B audits in prevision of legal proceeding is limited BY the fact that they can only found existing apparent elements in the environment. However, because those elements are existing and apparent, their correctness is guaranteed therefore their probative value should remain equal to the same evidence brought by hand. Using \B audits should have consequences on the admissibility of the evidence but should not impact its evidential value. 

A recent proposal from the european commission for an \textit{artificial intelligence act} \footnote{Proposal for a {REGULATION} {OF} {THE} {EUROPEAN} {PARLIAMENT} {AND} {OF} {THE} {COUNCIL} {LAYING} {DOWN} {HARMONISED} {RULES} {ON} {ARTIFICIAL} {INTELLIGENCE} ({ARTIFICIAL} {INTELLIGENCE} {ACT}) {AND} {AMENDING} {CERTAIN} {UNION} {LEGISLATIVE} {ACTS} {COM}/2021/206 final.} gives some insights about the future regulation strategy. This proposal wants to establish a certification system for high-risk AI system before they enter european market. Certification means precise and technical standards established publicly to help AI systems providers to comply. Furthermore, this proposal provides an obligation on the providers to create a "technical documentation\footnote{Article 11 of the proposal.}" in order to help the user. The technical elements included both in the legal standards and in the technical documentation could increase the situations where \B audit forms will be useful\footnote{At the time of writing, this is still a proposal which can be subject to numerous amendments before being voted by the european parliament. Besides, it is only aiming to high-risk AI systems, the extension of the certification logic to all algorithmic systems through others future regulations is not certain.}. Furthermore, this development logic of European standards could one day help improving the evidential value of the results from a certified audit algorithm. Indeed standards from european institutions, named in european regulations, have a determined legal value, stronger than scientific standards alone \cite{aMazeauResponsabiliteenjeuxnormalisation2019} \cite{waeyenbergeNormalisationTechniqueEurope2018}.

On the other hand, \SY audits provide interpolations that are most of the time hidden to the user. The objective of the \SY audits is to create a representation of the audited algorithm. Because by definition the audited algorithm is not open to the public, the correctness of the representation is not guaranteed. Every statement made on the surrogate is an assumption. As an illustration, we can take the example of someone who tries to assess the potential discrimination bias against women within a hiring decision system\cite{DBLP:journals/corr/abs-1906-09208}.

This potential inaccuracy, inherent to all statistic tools, will have consequences regarding the probative value given by the judge to the audit results. This new question has not been addressed by courts yet. Therefore it is not possible to propose an evidential value to \SY audits. It could become a strong evidence, like DNA in paternity test\footnote{On the importance of DNA in paternity test, see, among others :Cour de cassation, civile, Chambre civile 1, 25 septembre 2013, 12-24.588, Inédit, 2013. and Cour de cassation, civile, Chambre civile 1, 25 septembre 2013, 12-24.588, Inédit, 2013.}, or it could be a more contextual, secondary evidence as DNA in criminal proceedings\footnote{Because DNA does not provide all the elements necessary to establish guilt, its usefulness and utilization is actually limited. See Julie Leonhard, « La place de l’ADN dans le procès pénal », Cahiers Droit, Sciences \& Technologies, 9,  2019, 45-56. Also Olivier Pascal, “Empreintes génétiques au service de la justice. Arx Tarpeia Capitoli Proxima ou l’incertitude de la science” (2019) 9 Cahiers Droit, Sciences \& Technologies 39–43.}. 

It is crucial to realize that the use of an audit algorithm, with all the state of the art elements of scientific procedure followed, does not guarantee that the output will be considered as an important evidence from a court, compared to a testimony or others kind of evidence. Considering the importance of potential inaccuracy and postulates in audits from a user perspective will take time for lawmakers or judges. Meanwhile, the collaboration between legal experts and scientists is not only needed but also necessary considering the issues at stake.

\subsection{Auditor protection and the future of algorithms auditing}

\subsubsection{The absence of legal protection for external unauthorised auditors.}

Whistle-blower regulation initiatives emerge in some national law\footnote{In France, a legal protection is granted in 2016 through a law for transparency and against corruption.}. In 2019, 6 years after the Snowden scandal, the european union voted a directive on the protection of whistle-blowers\footnote{Directive ({EU}) 2019/1937 of the European Parliament and of the Council of 23 October 2019 on the protection of persons who report breaches of Union law.}. European member states must transposed the directive into their national law before December 17th, 2021. The 2019 european directive grants a legal protection for people who "acquired information on breaches in a work-related context [...]\footnote{Article 4 "\textit{personal scope} of the european directive"}. In the situation of audit algorithms, \textit{work-related context} means the whistle-blower had, has or is going to have a work relationship with the owner of the controversial audited algorithm. This new european legal protection forms a new milestone, but it seems to exclude some categories of people.

For instance, scientists or any external person who want to audit  a public algorithm and find out a violation to the law are left aside of this protection. National laws could extend this protection to third-party actors, but the scope of protection will differ between european member states\footnote{For instance, french regulation protects all individuals who identify and communicate a breach of security to the national cyber-security authority, showing the willingness to collaborate with \textit{white hats}.}. Researchers do not get a specific protection like journalist do. It means auditing algorithms without a proper authorisation could expose the auditors to legal consequences. 
Using \B or \SY audit forms is not only about improving the chances of going to court and obtaining a corrective sanction to the illegal situation by the audited algorithm. It is also a matter of reducing legal auditors responsibility once the audit is revealed, either through a court decision or a public article. 

\subsubsection{The future of online algorithms auditing}

The european directive "Directive services act proposal"\footnote{Proposal for a {REGULATION} {OF} {THE} {EUROPEAN} {PARLIAMENT} {AND} {OF} {THE} {COUNCIL} on a Single Market For Digital Services (Digital Services Act) and amending Directive 2000/31/{EC} {COM}/2020/825 final.} is bringing a fresh legal framework about the accountability of providers of intermediary services. First, the potential creation of a new regulatory entity called the "Digital Services Coordinator" could avoid the uncertainty about admissibility and probative value of auditing results. Indeed, every user of a platform\footnote{Called "recipient of the service" in the european proposal.} can lodge a complaint to their national entity if they find out about a violation of the regulation, \eg when a platform does not remove an inappropriate content even after being notified of its existence. Second potential improvement, this proposal contains an obligation for very large  platforms to be audited once a year by an external and independent auditor. Third and last improvement, the creation of the \textit{trusted flaggers} status\footnote{Article 19 of the digital services act proposal.}. Notification from those trusted flaggers will be processed "with priority and without delay" by platforms. This status will be awarded by the Digital Services Coordinator based on 1) the expertise and competence in detecting, identifying and notifying illegal content ; 2) its independence and the fact that it represents collective interests and 3) it focused on objective, fast and precise notifications. Without giving a legal protection to the auditors of online algorithms, this proposal gives a more central place to certain users regarding the content moderation of online platforms. This trusted flaggers status could be a first step to realize the importance of involving the civil society into the auditing of online platforms and algorithms, through the use of audit tools.

\section{Conclusion}

We have discussed that the outcome of audits --that we fit into two categories (\B and \SY)-- do not necessarily conforms to what is a proper building of a case. Multiple precautions must be carried out before and during the audit, in order for evidence to be considered so that they have an impact in practice. A central objective also being to avoid the auditor to be prosecuted.

There is an increasing amount of approaches to audit algorithms. In particular, there are some variants of \B, where real inputs are used to create synthetic ones, that will in turn be used against the algorithm \cite{mahmood2021black}. This makes the border between \B and \SY harder to grasp. An auditor must in consequence permanently monitor the advance of audit techniques, and the mutations of law.

Finally, we stress that the absence of the proof of bias in an audited algorithm does not mean that there are not issue in it. This echos the so called \textit{diesel gate}, and more technically the possible temptation for \textit{fairwashing} \cite{aivodji2019fairwashing}, where the audited algorithm can "sandbox" the auditor into an acceptable vision of its operation. In such a case, on more advanced regulation is awaited to have a fundamental impact on decision making algorithms.

\bibliographystyle{abbrv}
{
\bibliography{biblio}
}

\end{document}